# COVID-19 Diagnostics: Past, Present, and Future

Alexis Scholtz[1], Anuradha Ramoji[2,3], Anja Silge[2,3,4], Jakob R. Jansson[5], Ian G. de Moura[5], Jürgen Popp[2,3,4] Jakub P. Sram[5], Andrea M. Armani[1, 6, *]

[1]Department of Biomedical Engineering, University of Southern California, Los Angeles, California 90089, United States of America

[2]Institute of Physical Chemistry (IPC) and Abbe Center of Photonics, Helmholtzweg 4, 07743 Jena, Germany

[3]Leibniz Institute of Photonic Technology (IPHT) Jena, Member of the Leibniz Research Alliance - Leibniz Health Technologies, Albert-Einstein-Straße 9, 07745 Jena, Germany

[4]InfectoGnostics Research Campus Jena, Centre of Applied Research, Philosophenweg 7, D-07743 Jena, Germany

[5]Fulgent Genetics, Temple City, California 91780, United States of America

[6]Mork Family Department of Chemical Engineering, University of Southern California, Los Angeles, California 90089, United States of America

## ABSTRACT

In winter of 2020, SARS-CoV-2 emerged as a global threat, impacting not only health but also financial and political stability. To address the societal need for monitoring the spread of SARS-CoV-2, many existing diagnostic technologies were quickly adapted to detect SARS-CoV-2 RNA and antigens as well as the immune response and new testing strategies were developed to accelerate time-to-decision. In parallel, the infusion of research support accelerated the development of new spectroscopic methods. While these methods have significantly reduced the impact of SARS-CoV-2 on society when coupled with behavioral changes, they also lay the groundwork for a new generation of platform technologies. With several epidemics on the horizon, such as the rise of antibiotic-resistant bacteria, the ability to quickly pivot the target pathogen of this diagnostic toolset will continue to have an impact.

## INTRODUCTION

Over the past several decades, engineers and medical practitioners have worked symbiotically to develop tools and instruments to accelerate disease diagnosis[1–3]. This concerted effort enabled the transition from time-consuming imaging and cell culture-based diagnostics to rapid high throughput genetic and protein analysis. In the past few years, these methods have been further improved by the integration of robotic sample handling and preparation and by data analytics based on artificial intelligence (AI). As platform technologies, these advances have cut across all fields in medicine, improving patient care.

While many of these innovations were motivated by cancer and heart disease, with the onset of COVID-19, several technologies were quickly pivoted to address this worldwide pandemic[4–11]. Additionally, the clear global need motivated many academic research teams to shift their focus from basic science to more applied research. However, the medical and financial requirements between these two classes of disease, from chronic conditions to acute infection, are very different. Therefore, it quickly became evident that some assays were more ideally suited for this shift.

In this perspective, we will provide an overview of some key metrics when evaluating the utility of a given sensor for a diagnostic application. We will then discuss several examples of commercialized systems that successfully pivoted from their original purpose and have now made a significant impact in controlling the spread of COVID-19. Lastly, we will present emerging optical diagnostic methods suitable for SARS-CoV-2 detection that are currently under development.

## SENSOR METRICS

The two primary characteristics when evaluating a diagnostic sensor are sensitivity and specificity. Sensitivity is the true positive rate, or in the case of a diagnostic, the proportion of sick patients who test “positive”[12]. The specificity is the proportion of healthy people who test “negative,” and is therefore the true negative rate. Depending on the exact diagnostic test, many factors can contribute to these metrics, including the detection mechanism, the sample type, and the sample preparation requirements[13].

The false positive rate, also known as a false alarm or type I error, can be derived from the sensitivity as the rate of healthy patients who test “positive”. The false negative rate, also known as a miss or type II error, can be derived from the specificity as the rate of sick patients who test “negative”[12]. Ideally, a diagnostic would have both high sensitivity and high specificity, but, as in any binary classification, there is a trade-off between these two metrics that can be set depending on the application. For a screening test, such as screening a population for COVID-19 to isolate infected individuals, it is advantageous to set a higher sensitivity threshold. This strategy allows the test to catch more of the cases, improving the overall effectiveness of the quarantine process. Figure 1 demonstrates this trade-off from the standpoint of changing the threshold that determines which cases are considered “positive” versus “negative”.

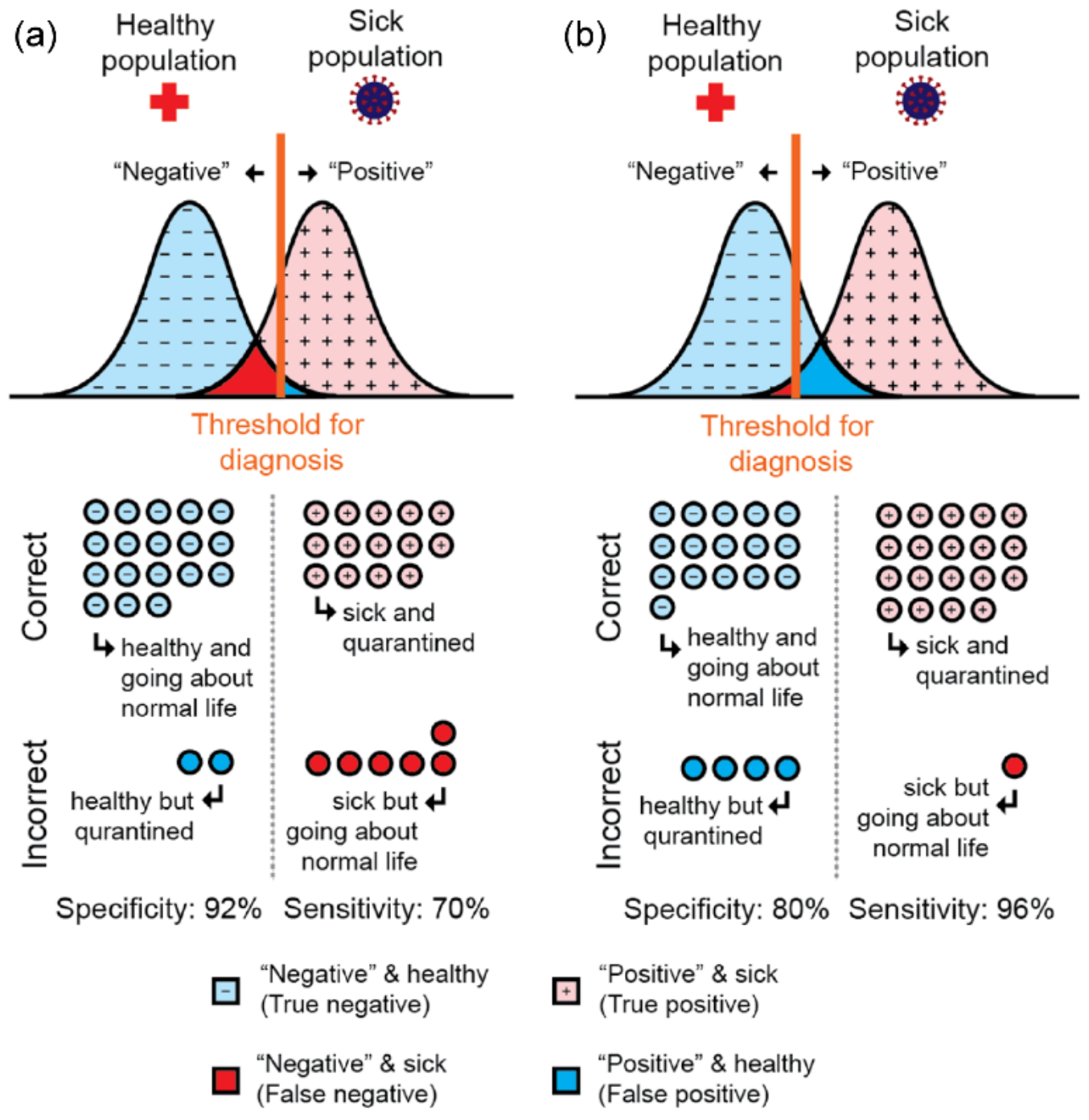

Figure 1: An illustration of the trade-off between sensitivity and specificity. (a) A higher threshold, depicted by the orange line, for diagnosing a "positive" case results in a higher specificity but lower sensitivity, leading to a higher false negative rate. (b) A lower threshold for diagnosing a "positive" case results in a higher sensitivity but lower specificity, leading to a higher false positive rate. While this approach will result in a larger number of healthy individuals being unnecessarily quarantined, it will correctly diagnose a larger percentage of the infected population. Therefore, this approach may be the preferred strategy for a screening test for a highly infectious agent.

An important factor in determining the sensitivity of a test is the sample collection and purification. Methods recommended by the United States Center for Disease Control and Prevention (CDC) include nasal swabs, throat swabs, saliva collection, and blood samples[10,14]. Studies have suggested that of these methods, performing throat swabs alone may lead to reduced sensitivity relative to the other techniques, but there is not yet a consensus on the optimal specimen collection method. Within each of these samples, SARS-CoV-2 is detected either directly, through the presence of viral RNA[15] or antigenic macromolecules located on or within the virus[16], or indirectly, by detecting the patient's immune response (i.e. antibodies) to the virus[17–19]. It is important to note that blood-based serological testing typically detects the antibodies that are produced after the initial infection. Therefore, these antibody-based strategies are more applicable for testing for past illness rather than diagnosing active infection[20].

Another important metric is the limit of detection (LOD), or the smallest concentration of analyte in a sample that can be detected[21,22]. A lower LOD will increase sensitivity since the test can detect lower levels of analytes, as may be the case shortly after infection. As already discussed, increasing sensitivity is important. However, this goal should be pursued within the relevant concentration ranges of the specific disease or illness. Additionally, it is important to optimize the sensor response to occur within the linear working range of the sensor[21]. This allows the sensing signal to be directly correlated to the diagnostic biomarker.

Both the LOD and the linear working range are dependent on the test type and instrumentation rather than the sample collection method. Therefore, optimization of the detection or sensor technique and the chemical assay and reagents are the primary limiting factors and one of the main drivers of innovation in the field. For example, high-throughput laboratory equipment can detect quantities below 10 copies of SARS-CoV-2 RNA/mL while point-of-care (POC) instruments had LODs of >100 copies/mL[4,23,24]. However, the precise LOD of the POC instruments varies greatly among the different manufacturers.

Because the actual SARS-CoV-2 viral loads range from <100 copies/mL up to $>10^6$ copies/mL[10,25–27], with peak viral load occurring a few days after symptoms show, the ability to detect a viral load early is system-dependent. For comparison, these values are similar to two other viral diseases whose viral loads are commonly quantified through similar methods: human immunodeficiency virus and hepatitis C virus[28–32].

Other important metrics include reproducibility, the ability of a diagnostic to produce the same results when the same sample is tested repeatedly, which is important in establishing the reliability of diagnostic tests[12]. Because many immunoassays and viral diagnostic methods had already demonstrated reproducibility for other diseases and pathogens, they were ideal candidates for COVID-19 diagnostics.

## METHODS MAKING AN IMPACT

Typically translating a technology from a research setting to use in patient care takes years. However, this process was accelerated by changes in regulatory process and increases in government funding during the spring and summer of 2020 to address the imminent societal threat posed by COVID-19 (Figure 2). This support fueled the development of diagnostics in both academic and industry research labs.

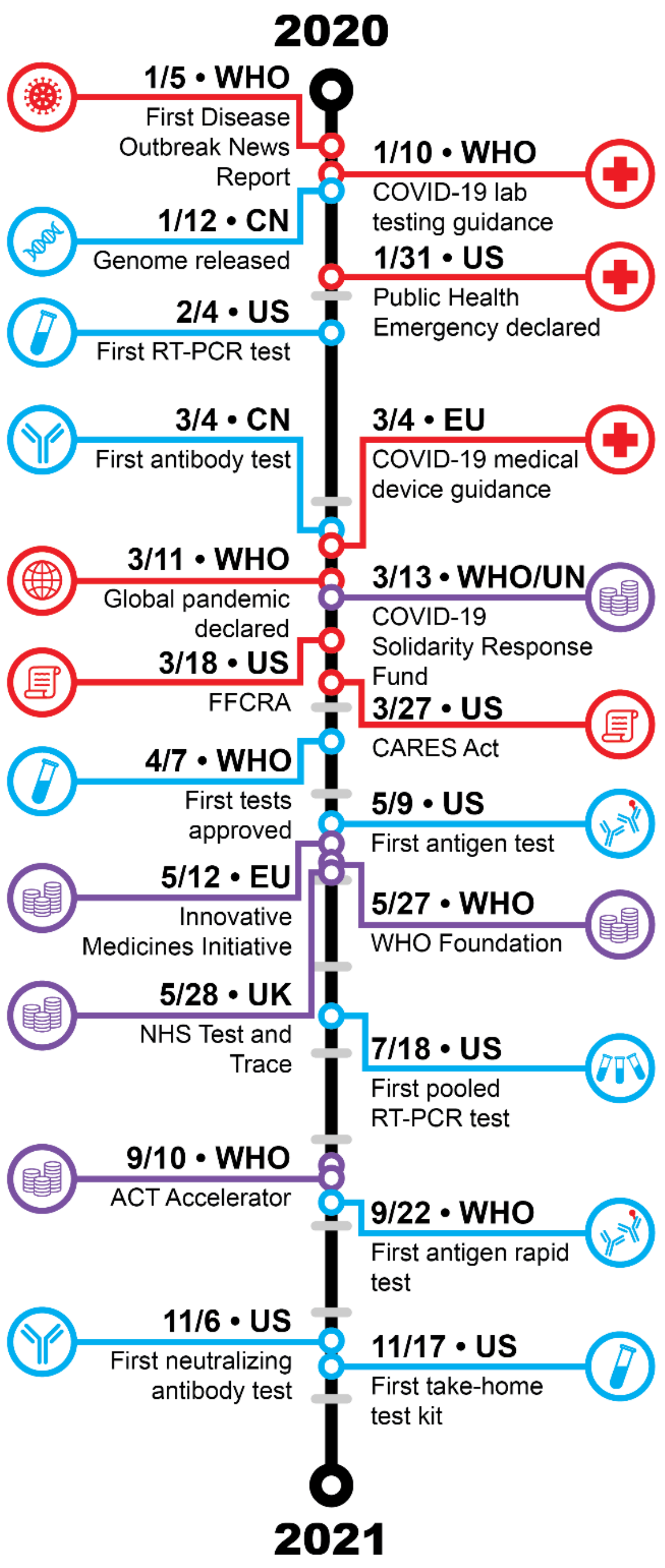

Figure 2: Timeline of major funding (purple), regulatory and institutional (red), and scientific developments (blue) during the first year of the COVID-19 pandemic. Country and organization abbreviations: WHO, World Health Organization; CN, China; EU, European Union; UK, United Kingdom; UN, United Nations; US, United States. [33–41]

At the onset of the COVID-19 pandemic, both diagnostic and antibody testing methods already existed for other coronaviruses, such as SARS and MERS. However, developing, validating, and mass-producing tests for the detection of SARS-CoV-2 in symptomatic and asymptomatic individuals presented a significant challenge which required largescale international and multi-stakeholder collaborations to overcome.

On January 5, 2020, the World Health Organization (WHO) issued its first Disease Outbreak News Report, officially acknowledging the pneumonia cases of unknown origin in Wuhan. Several days later, after it was discovered that these cases were caused by the novel coronavirus that would come to be known as SARS-CoV-2, they published new laboratory testing guidelines for novel coronavirus infections[33], and the first SARS-CoV-2 genome sequence was released by China. This sequence of events assisted numerous governments and laboratories with their development of RT-PCR tests targeting different gene sequences unique to SARS-CoV-2.

In March 2020, the European Union (EU) announced their guidelines for *in vitro* diagnostic medical devices in the COVID-19 context, which further helped to standardize COVID-19 testing in the EU. By this time, the WHO declared the COVID-19 crisis to be a global pandemic, and, with the United Nations (UN) established the COVID-19 Solidarity Response Fund to start leveraging private donations. The WHO went on to establish the WHO Foundation and the Access to COVID-19 Tools (ACT) Accelerator, both of which provide grants to research scientists. These unique collaborations between academic, governmental, and private institutions, brought about by the unifying fight against COVID-19, have sparked rapid biomedical innovations worldwide, including in the diagnostic realm. Within Europe, these efforts were aided by the Innovative Medicines Initiative's call for rapid, POC COVID-19 diagnostics and the United Kingdom's (UK) National Health Service (NHS) Test and Trace program, whose goals included bringing mass COVID-19 diagnostic testing to the public.

Meanwhile, the United States (US) had declared a public health emergency in late January 2020, which allowed the Food and Drug Administration (FDA) to begin issuing Emergency Use Authorizations (EUAs) for COVID-19 tests, offering an expedited regulatory pathway to facilitate rapid development for the emerging viral threat[34]. Subsequent American funding initiatives passed in the following months, including the Families First Coronavirus Response Act (FFCRA)[37] and Coronavirus Aid, Relief, and Economic Security (CARES) Act[36], provided additional incentive for the development of diagnostic tests based on nucleic acids, antigens[38], and antibodies[40].

As a result of these investments, numerous COVID-19 diagnostic strategies were successfully developed and deployed for use in clinical settings globally. They can be categorized as: (1) Nucleic Acid Amplification Tests (NAATs), (2) Next Generation Sequencing (NGS) tests, (3) Antigen tests, and (4) Antibody tests. A summary of current testing methodologies can be found in Table 1[42].

| **Test** | **Test Type** | **Sample** | **TAT** | **Strength** | **Weakness** | **Use Case** |
|---|---|---|---|---|---|---|
| RT-PCR | NAAT | Nasal, Saliva | 24h-72h | High specificity and sensitivity | Longer TAT than some alternatives | Gold standard for SARS-CoV-2 diagnostic testing |
| Rapid RT-PCR | NAAT | Nasal, Saliva | >1h | Fast TAT | Low volume | On-Site Testing |
| RT-Lamp | NAAT | Nasal, Saliva | <24h | High specificity and simplicity | Limited access to affordable reagents | On-Site Testing |
| Pooled RT-PCR | NAAT | Nasal, Saliva | 24h-72h | Economical | Repeat test required if positive result is returned | Surveillance Testing |
| NGS | NGS | Nasal, Saliva or Blood | <72h | Can identify the specific SARS-CoV-2 strain/variant with high specificity and sensitivity | High cost and longer TAT | Identify variants of concern and help validate testing methodologies |
| Antigen | Antigen | Nasal | <30min | Fast TAT | Low specificity and sensitivity | On-Site Testing |
| Antibody | Antibody | Blood | 24h-72h | Identifies a previous SARS-CoV-2 infection | Cannot identify a current SARS-CoV-2 infection | Identify immunological response to SARS-CoV-2 (past infection) |

Table 1: Overview of testing methodologies currently in clinical use.

**Nucleic Acid Amplification Testing**

The majority of COVID-19 nucleic acid tests rely on first converting viral ribonucleic acid (RNA) to deoxyribonucleic acid (DNA) using reverse transcription (RT) and then subsequently amplifying the concentration of the DNA using polymerase chain reaction (PCR)[15,43,44]. The different types of nucleic acid-based tests are differentiated by the number of steps used and the specific approaches used to obtain the final DNA read-out. However, one limitation with all RT-PCR tests is that they are not able to detect previous infections.

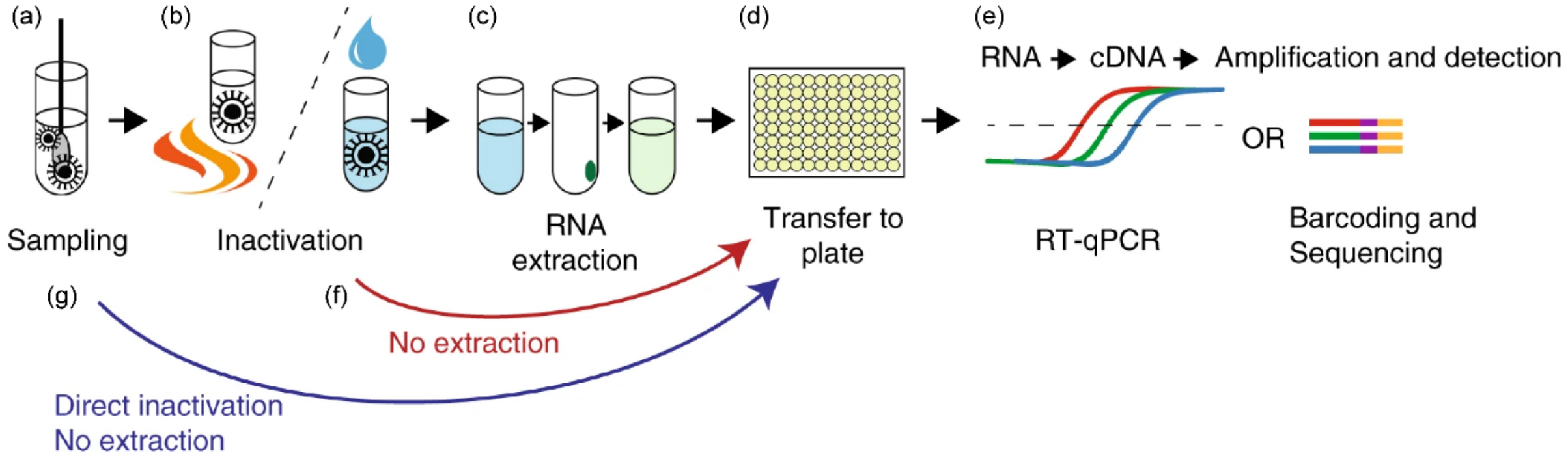

Figure 3: Overview of the RT-PCR process. (a) First, sample specimens are taken from the patient. (b) Then the sample is treated to inactivate the SARS-CoV-2 virus, if present. (c) Next, RNA extraction is performed on the sample. (d) Samples are pipetted into a reaction plate, and (e) the RNA is analyzed via RT-qPCR. If any viral RNA is detected, the test result will be positive. Alternative methods also exist which deploy (g) direct inactivation and (f) omit the extraction step. Adapted from (15). Distributed under a Creative Commons Attribution NonCommercial License 4.0 (CC BY-NC) http://creativecommons.org/licenses/by-nc/4.0/

SARS-CoV-2 PCR uses a SARS-CoV-2 primer and probe set to recognize and copy target regions in the SARS-CoV-2 genome (Figure 3). A second primer and probe set that detects human RNase P (RP) can be used as an internal control or standard. Amplification of the SARS-CoV-2 markers and control targets are then performed using PCR methods. Initial tests performed these steps iteratively, first converting RNA to complementary DNA (cDNA) and then amplifying and detecting that cDNA using fluorescence or barcoding techniques. However, with the optimization of testing conditions and sample pre-processing, it was possible to design a test that could be performed in a single vial.

This Rapid RT-qPCR test has several advantages, but most importantly, it reduces the required reagent, thereby diminishing resource limitations on test capacities, facilitating a faster Turnaround Time (TAT), improving the effectiveness of quarantine procedures, and allowing testing at POC. Notably, RT-PCR can provide diagnosis for both symptomatic and asymptomatic patients.

This advance was accomplished in two ways: (1) several companies developed a hand-held thermal cycler that allows for the RT-qPCR test to be run outside of a lab; (2) other companies like Abbott developed an isothermal RT-qPCR test utilizing a DNA polymerase that can melt and synthesize DNA at the same temperature, removing the need for laboratory-bound equipment[23,45–47]. In the latter, the reactions for PCR are performed on-site in different test tubes or in a lateral flow assay at room temperature.

For these systems, the enabling technology advances can be broadly categorized into two classes: 1) accelerating or simplifying the PCR amplification process or 2) simplifying the DNA detection. One of the primary approaches for simplifying the PCR was using microfluidics with an integrated thermocycler and robotic sample handling. To reduce the complexity of DNA detection, companies integrated spectroscopy and fluorometry detection systems optimized for the relevant probe wavelengths. This approach reduced sampling handling requirements and accelerated detection.

One variation of RT-qPCR which began to gain popularity over the past year was loop-mediated isothermal amplification (LAMP) technology[48]. Similar to some RT-qPCR tests, RT-LAMP utilizes a DNA polymerase that isothermally melts and synthesizes DNA. It also includes an RNA transcriptase so that reverse transcription and DNA amplification can occur simultaneously. If the unique viral gene targeted by the assay exists in the sample, DNA amplification will occur. This increase in DNA will decrease the pH of the sample, causing a color change that can be read with a cell phone, without the need for laboratory processing. However, at this time, the reagents are unreliable and difficult to acquire, limiting its use[49].

### Next Generation Sequencing

One disadvantage of the RT-qPCR approaches is that *a priori* knowledge of the sequence is required to create the necessary primers. As new variants of SARS-CoV-2 began to emerge, the need for broader screening technology that does not rely on prior knowledge of the variants' genetic sequence came to the forefront. Next Generation Sequencing (NGS) Testing allows surveillance, detection, and characterization of different SARS-CoV-2 strains[50].

There are two methods for sequencing SARS-CoV-2: (1) shotgun metagenomics, and (2) target enrichment. Shotgun metagenomics sequences all DNA/RNA in a sample without prior knowledge of the sources of the genetic material. This method allows for high-throughput sequencing and identification of co-infections to help tailor treatment plans. NGS relies on breaking DNA or RNA into short fragments, sequencing, and then re-assembling to determine the entire genomic sequence. If a target enrichment strategy is used, only specific genomic regions of interest in a sample are sequenced, allowing for faster sequencing and improved accuracy and specificity. Target enrichment allows scientists to generate targeted NGS libraries, and it is an effective method for developing panels that can target multiple pathogens at once (Figure 4). Numerous photonic technologies have been used to improve the sensitivity of NGS measurements, including bead-based methods and optical absorption and fluorescent techniques.

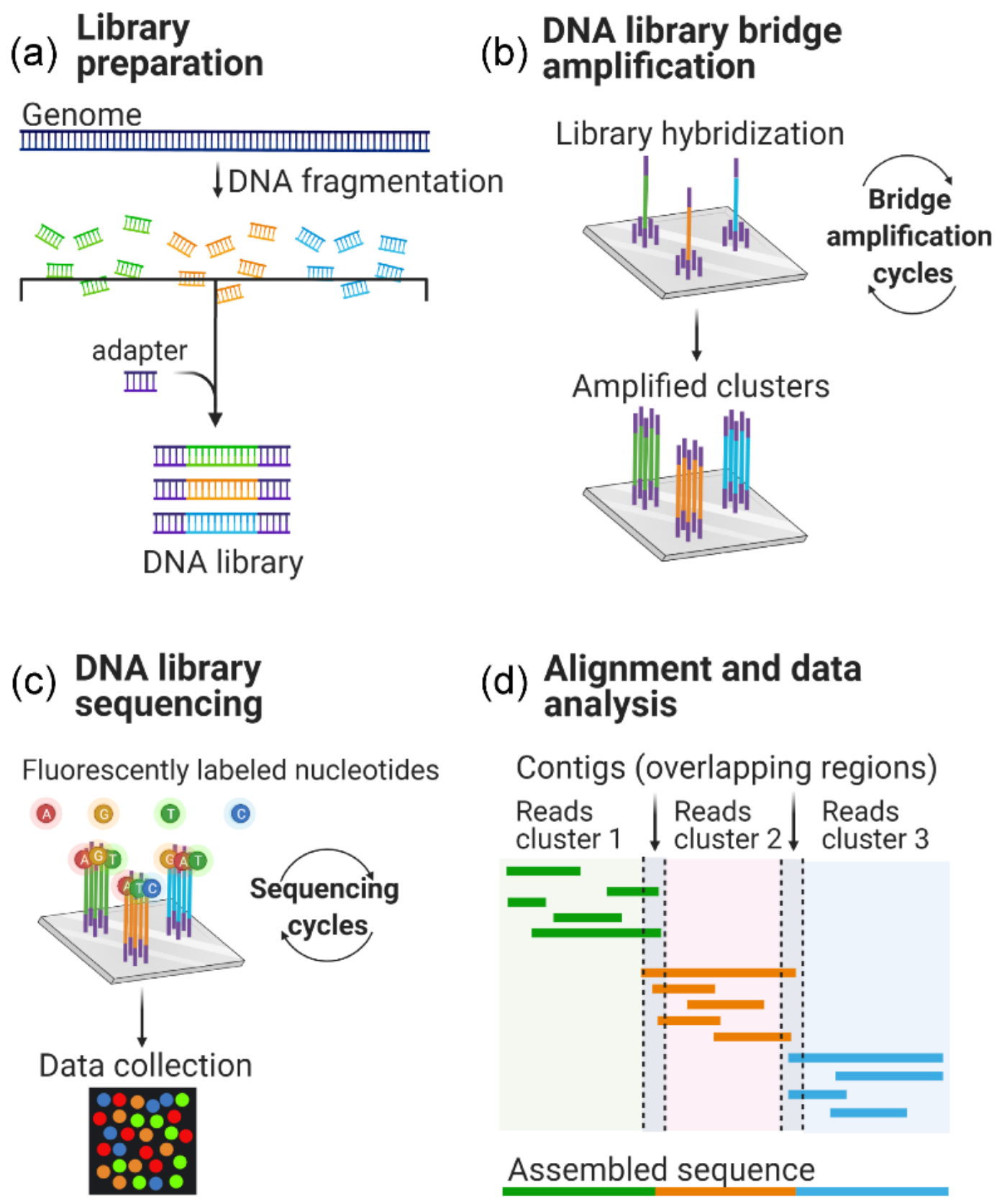

Figure 4: Next Generation Sequencing process. (a) The NGS Library is prepared by fragmenting the target DNA and hybridizing the fragments with specialized adapters. (b) The NGS Library is then loaded onto the sequencing plate and amplified. (c) Fluorescently labeled nucleotides are added to the plate and the nucleotide binding patterns are recorded to sequence the NGS library. (d) Finally, these sequences are aligned and the assembled sequence is analyzed through a bioinformatics pipeline; bioinformatics scientists compile the data and reconstruct the complete NGS library. Created with BioRender.com.

## Antigen Testing

The rapid antigen COVID-19 test is an immunoassay that detects SARS-CoV-2 viral proteins in respiratory samples[51]. The test utilizes a sandwich assay where the sample is mixed with a solution containing two antibodies. Both antibodies will bind to SARS-CoV-2 viral proteins, if present. The first antibody will anchor the viral protein to a detection sheet, and the second will tag the viral protein for detection, typically via fluorescence. Because the test requires a SARS-CoV-2-specific antibody pair, the antigen tests took a longer time to develop and validate than the RT-PCR tests[52].

While antigen testing does prove effective in identifying likely infectious individuals, or individuals in a population with a high concentration of positive cases, RT-PCR tests remain the gold standard due to their higher sensitivity, specificity, and scalability. However, demand for antigen testing has recently increased due to its affordability, rapid TAT, and ability to be administered at the POC without the need for complex instrumentation.

**Antibody Testing**

In contrast to antigen tests, the COVID-19 antibody test is an immunoassay that detects IgG and IgM antibodies generated by the immune system in response to SARS-CoV-2[53]. Notably, unlike previously discussed tests, antibody-based tests are performed using patient blood samples, many at the POC, and the results can be delivered within 30 minutes.

As shown in Figure 5, antibody-based tests utilize a sandwich assay where the sample flows across a substrate labelled with several capture IgG and IgM antibodies specific to SARS-CoV-2 as well as a control[17–20,53]. For a fluorescent-based test, the antibodies are labelled with fluorescent tags that allow for detection of the SARS-CoV-2 antibodies. This approach allows for the same instrumentation as a standard enzyme-linked immunosorbent assay (ELISA) to be used to read-out the result. Alternatively, metal nanoparticle labelling indicators are subsequently added. These are typically gold nanoparticles, which appear red due to the surface plasmonic resonance (SPR) of the particles. Using these metal nanoparticles and SPR instead of a conventional fluorophore reduces the sensor's susceptibility to photo-degradation and increases its shelf-life. Additionally, in many cases, the result can be determined "by eye" or visually.

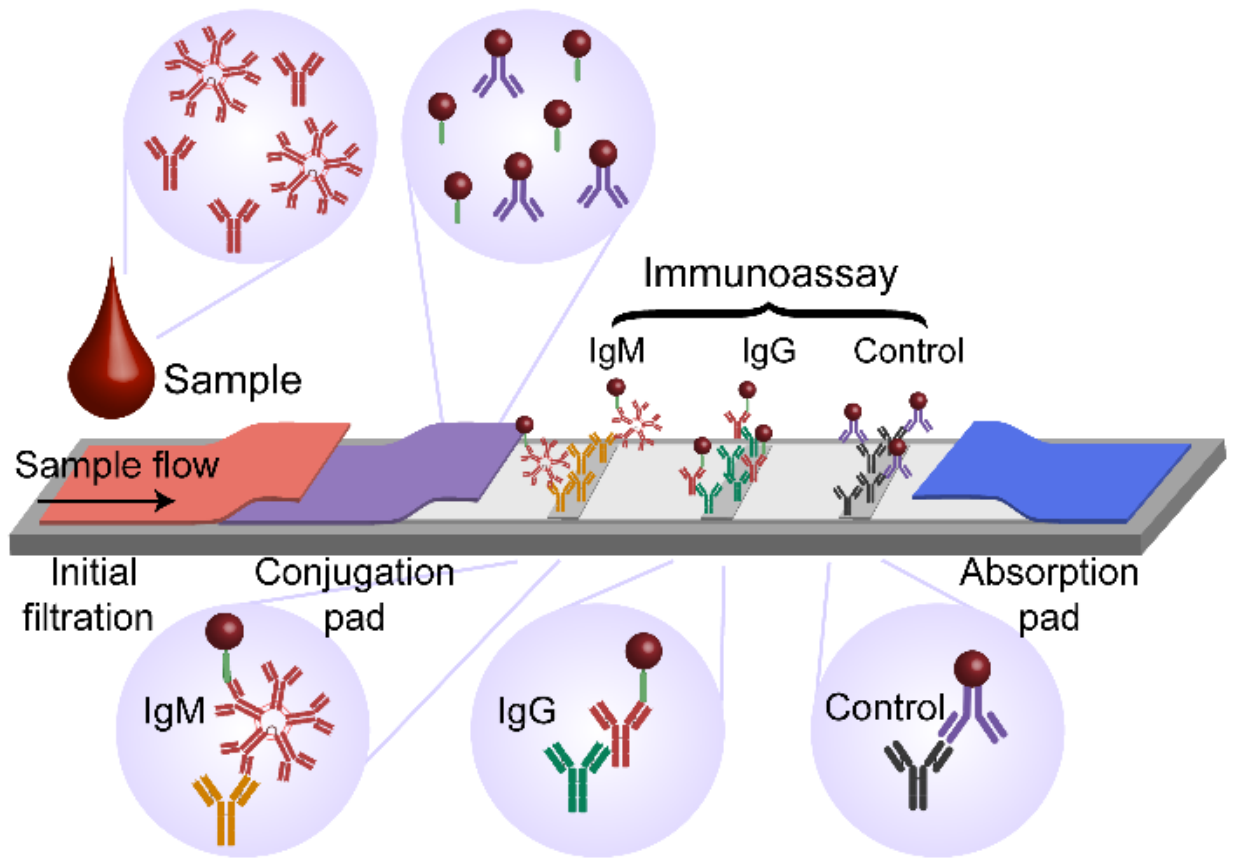

Figure 5: Schematic of a rapid diagnostic test (RDT). (a) The RDT has two diagnostic lanes and a control lane. The conjugation pad contains a SARS-CoV-2 antigen and a control antibody, both labelled with a metal nanoparticle. The sample is wicked across the conjugation pad and then across all three lanes. (b) If a strip changes color, it indicates that the antibody is present. Adapted from (11).

However, the specificity and sensitivity of antibody-based tests are dramatically reduced when compared with nucleic acid-based tests. Additionally, as mentioned, the utility of an antibody test for detecting an active infection is moderate to low because the test is detecting the immune response that begins days after initial infection. The ideal use for these tests is in identifying asymptomatic individuals who were unknowingly infected. This information can help determine the percentage of the population already infected by SARS-CoV-2. Alternatively, antibody testing for symptomatic cases in which the patient has previously had COVID-19 may shed light on how the body responds to a second infection by SARS-CoV-2.

Another difference between immunological methods and nucleic acid techniques is the ability to quantify an infection. RT-qPCR testing can quantitatively determine the viral load. With a better understanding of COVID-19, this capability could provide physicians with an indicator of when the infection occurred, allowing improved contact tracing and informing treatment strategies. However, we do not currently have the data necessary to draw these conclusions. Therefore, tests typically report only the presence or absence of SARS-CoV-2, based on a defined PCR cycle threshold.

**Pooled Testing**

Given the scale of the pandemic and its impact on manufacturing and supply chains, the demand for PCR-based tests often outpaced capabilities. This gave rise to the implementation of novel testing approaches. One of the most broadly used strategies, pooled testing, allows for the monitoring of entire populations for outbreaks while simultaneously reducing the cost and the number of tests required[54–56]. It is important to note that pooled testing is not dependent on the diagnostic method used; however, pooled RT-qPCR was the most popular due to the availability and advantages of RT-qPCR tests.

Pooled RT-qPCR testing follows the same protocols as the standard RT-qPCR test (Figure 6), except that samples from multiple patients are combined and tested via RT-qPCR simultaneously. If SARS-CoV-2 is detected in the pooled sample, individual samples must be tested separately to identify the source of the positive result. Thus, to minimize repeated testing, it is more suitable when the probability of a positive result is low, such as when disease prevalence in the community is lower. The concept of pooled testing as a surveillance and monitoring strategy has broad applications beyond the present COVID-19 pandemic. This strategy could be used in the future as an approach to contain highly contagious pathogens. Given the ability of RT-qPCR to detect both symptomatic and asymptomatic cases due to its high sensitivity and the ability to tailor this testing strategy to detect the RNA or DNA of any pathogen, this testing strategy could find many applications in the future.

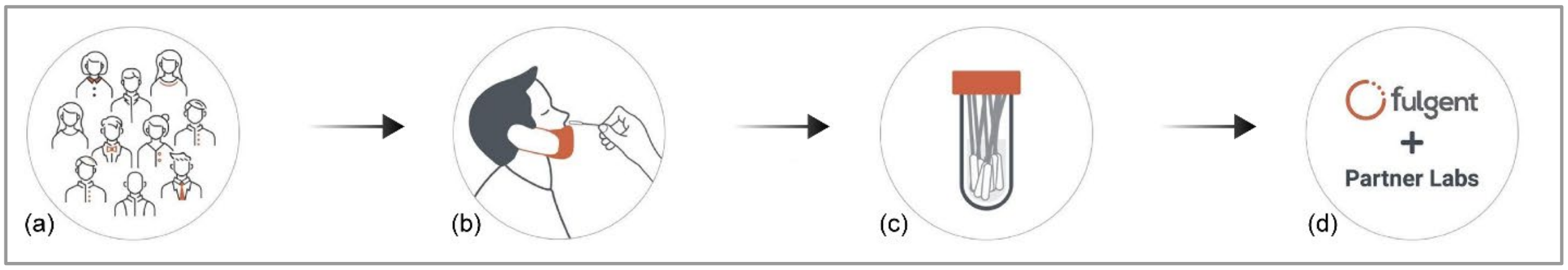

Figure 6: An example of a pooled testing strategy used in Los Angeles, California USA. Logo credit: Fulgent Genetics.

**EMERGING METHODS**

While previously commercialized systems are meeting the immediate needs of the medical community, additional optical methods based on detecting these same viral and immunological biomarkers are exploring ways to further increase sensitivity, decrease cost, and improve portability. These sensor systems include evanescent field sensors[6–8,51,53,57–60] and nanoparticle-enhanced imaging[61] as well as nanomaterial-based techniques[9]. The general design philosophy underlying all these emerging techniques is to improve the sensitivity of the detector. In doing so, the overall system performance will improve.

However, because these newer systems rely on detecting RNA, antigens, or antibodies, they face many of the same limitations regarding reagent shortages for PCR and specificity barriers for immunoassays. Thus, there is a strong motivation to pursue the development of methods that leverage completely new approaches for identification of the target. Additionally, there is an increased awareness of the importance of developing techniques that can non-destructively analyze a sample, allowing for multi-modal analysis or confirmation of a result.

Two optical-based methods that can meet both these needs are Raman spectroscopy and imaging (i.e. x-ray or computed tomography) of lung tissue. While Raman spectroscopy is still in the developmental phase, imaging has already begun to appear in clinical settings[59,62]. However, the use of imaging results as a primary diagnostic is unclear because there are many potential confounds. Therefore, the applications of imaging in COVID-19 are primarily used as either a pre-screen or for post-infection monitoring[63]. Additionally, performing imaging and the subsequent data analysis is very time-consuming, and high throughput imaging is unlikely to appear in the near future. Given these complications, we will focus on emerging Raman spectroscopy methods.

**Raman Spectroscopy**

One strategy that is being pursued is based on the detection of the Raman signature of molecules. The advantages of Raman spectroscopy are its versatility, high molecular specificity, and compatibility with small sample volumes. Additionally, Raman spectroscopy is non-destructive to samples. Unlike methods that rely on probe molecules for target identification, Raman spectroscopy leverages the vibrational signature of molecules within the sample that is inherent and unique to the infection of interest. While Raman spectroscopy is a well-established method, developing a diagnostic based on Raman spectra has been out of reach due to the complexity of deconvoluting the signature of the target of interest from the complex background of the biological matrix. However, with recent advances in data science, this hurdle has been overcome, and now, Raman spectral analysis is enabling richer data sets to be acquired. For example, the Raman spectrum can be used to determine the chemical composition of the target or subtle chemical changes within the sample under investigation[64–66]. Alternatively, in the context of COVID-19, the system can be configured to serve as a diagnostic[67,68].

A wide range of biological sample types including blood, saliva, and urine have been analyzed using Raman spectroscopy systems. The schematic in Figure 7 provides a guide for the Raman-spectroscopy based approaches that have already passed the proof-of-concept phase for characterization of bacterial infections and are currently being expanded to COVID-19 diagnostics[59,65,67,69–72]. Notably, on-site sample preparation strategies for chip- or cartridge-based miniaturized setups, compact Raman instrumentation, and data analysis pipelines for diagnostic information are interconnected and coordinated with each other. Furthermore, since Raman spectroscopy is a non-destructive method, it can be easily combined or confirmed with other methods including the EUA-approved RT-qPCR, making it a very attractive strategy.

The most promising isolation technique for a virus-sensing Raman platform is to capture and enrich particles via biochemical interactions. Capture probes, such as antibodies, bioreceptors, or specific surface proteins, some of which have been developed for COVID-19 diagnostics as previously discussed, have the ability to recognize and concentrate microbial targets on surfaces for Raman spectroscopic identification[73,74]. The detection concept aims at a qualitative detection of intact virus particles as an infectious entity. This prior work laid the foundation for this technology to be used to detect SARS-CoV-2 infections.

In an initial study using healthy and doped patient samples, SARS-CoV-2 was successfully detected[75]. However, the difference between the two samples is very subtle. To address this limitation, machine learning was incorporated into the spectral analysis. One example research effort is the development of the RNA Virus Detector (RVD) Graphical User Interface (GUI). Using this software, a detection accuracy of 91.6% was demonstrated[75]. While developed for COVID-19, this automated spectral analysis tool could be easily adapted to help manage future epidemics and pandemics.

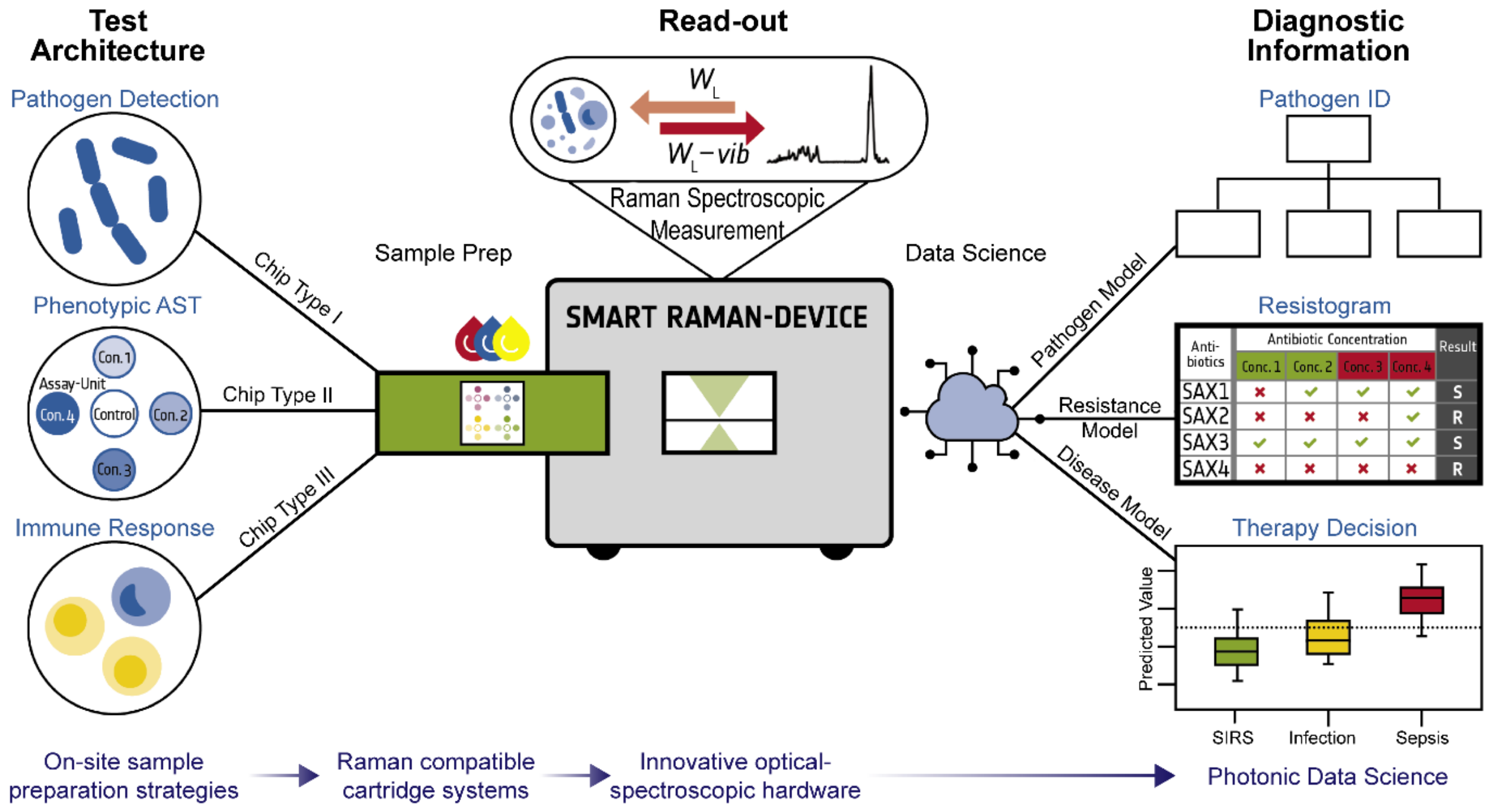

Figure 7: Schematic showing how on-site sample preparation strategies, Raman microspectroscopy and photonic data science can be combined to form innovative clinical diagnostics. The test architecture describes clinical issues that can be processed with the help of a bioanalytical chip platform. Miniaturized cartridges for sample preparation or assay units equipped with an optical window for laser access fit into compact spectroscopic hardware. Spectral information is translated into diagnostic information by photonic data science concepts.

A classic approach to increase the sensitivity of Raman-based detection is to use surface-enhanced Raman spectroscopy (SERS)[76]. In combination with an AI-based data analysis technique, this method allowed reliable detection of the SARS-CoV-2 virus particle in saliva samples of COVID-19-positive patients (Figure 8)[68]. The screening requires only a few minutes and demonstrated detection of SARS-CoV-2 with a sensitivity of over 90%.

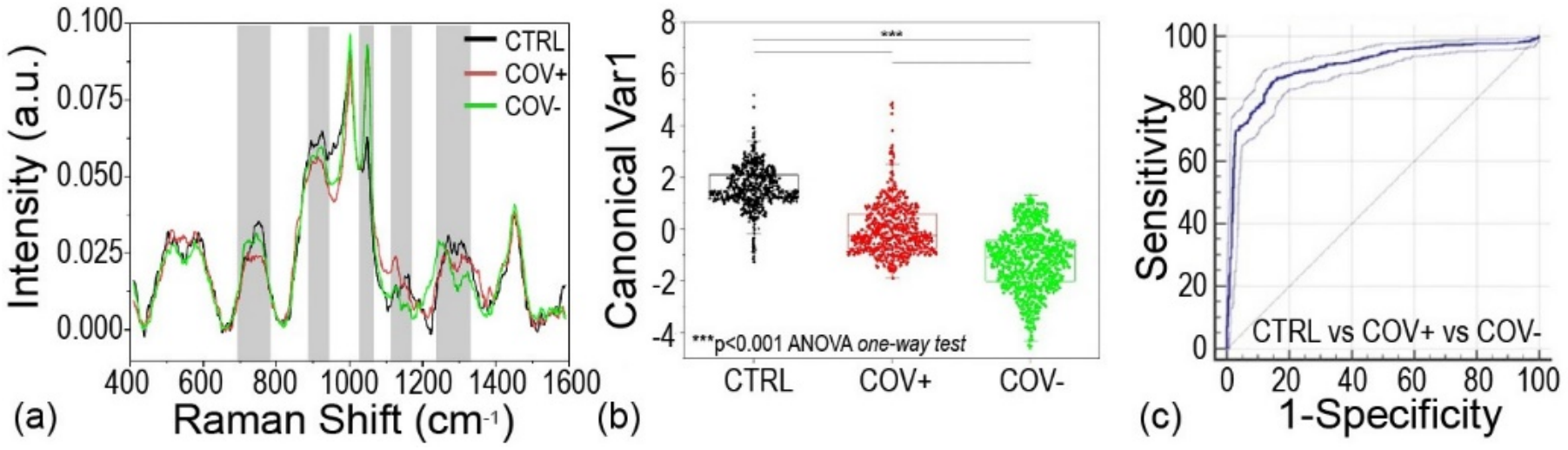

Figure 8: Combination of AI and Raman spectroscopy to detect SARS-CoV-2. (a) Spectral signatures of three samples and identification of key spectral features. (b) Correlation of sample type with the feature. (c) Diagnostic sensitivity vs. specificity for the variable from part (b). Adapted from (68). 

In summary, Raman spectroscopic identification of pathogens can be realized in less than 10 minutes, if the sample preparation for the respective specimen type and the reference database for the targets are established and optimized. The primary bottleneck is the isolation of the pathogen, which can be challenging if the sample matrix is complex (e.g. blood) or when a small number of cells is expected in a relatively large sample volume (e.g. when 1 cell has to be found in a sample volume of 100 ml)[42].

Just as RT-qPCR and antigen testing can be used to detect the presence of viral particles themselves and antibody-based serological testing can be used to detect immune responses, the examination of not just the pathogens themselves but also the immune response in body fluids by Raman spectroscopy has been an area of active research, and many of these efforts shifted focus last year. For example, Yin et al. demonstrated the detection of the immune response towards SARS-CoV-2 infection by screening the Raman spectral profile of the serum[67]. Therefore, the immune response to infectious diseases can be detected by Raman spectroscopy, provided that the heterogeneity of immune cells and specific Raman techniques for studies of body fluids are considered together with objective and robust statistical analysis methods.

To improve the robustness of the sensor result, researchers are developing systems that employ multiple detection strategies in parallel. It is important to note that this type of multi-modal system employing Raman spectroscopy has yet to be demonstrated for COVID-19; however, successful development of such a system for detecting other infectious diseases has already been demonstrated. For example, within the diagnostic platform described in Figure 7, pathogen identification can be combined with antibiotic sensitivity testing, allowing the susceptibility of a bacteria to an antibiotic to be identified or confirmed using two different technologies simultaneously[1,77,78]. Further, the biochemical signature activation profile of immune cells can be used to reliably diagnose infection and sepsis[79].

Biomedical Raman spectroscopy has evolved on both the technology and application side, indicating the extensive capabilities of this method. Nevertheless, direct implementation to translational applications has been rather cumbersome. In spite of the advantages of Raman spectroscopy, there are still several hurdles that must be overcome before it will be integrated into the clinical workflow. First, the sample preparation and measurements rely on highly trained scientists. Second, the measurement methods and data analysis are very labor intensive. While the RVD GUI is a significant first step in automating the analysis, until there are standard protocols, this method will be unattractive to clinicians. However, the research community is actively working on solving these problems, and several solutions are outlined in Figure 7[80].

## CONCLUSIONS AND FUTURE OUTLOOK

The experience with the COVID-19 testing response has demonstrated the value of providing alternative, accelerated regulatory pathways to motivate and catalyze the development of assays. The ability of labs to quickly validate pre-existing methodologies for the purpose of SARS-CoV-2 detection and the validation of less established methodologies, like RT-LAMP, resulted in the development of tools that might have otherwise taken years to develop. In addition to sensor technology, advances in signal processing and statistical methods have enabled the field of photonic diagnostics to move forward. The impact was most clearly observed in the implementation of pooled testing and in the use of automated data analysis programs.

While existing technologies took advantage of prior development to allow for quick adaptations to SARS-CoV-2 detection, they are still reliant on biochemical reagents, a resource that can limit test availability and accessibility, and biomarker identification, which can limit test specificity. Moving forward, optical technologies are being explored as a method that avoids these limitations; a notable example is Raman spectroscopy, which utilizes the inherent spectral fingerprints of pathogens. These optical methods provide useful advantages including: the tests are no longer reliant on the same reagents and are therefore less resource limited; they can be label-free and therefore don't require the same *a priori* knowledge of the pathogen's genetic sequences to create probes or the development of complementary antibodies; and these methods are non-destructive, which allows the test to be combined with other analysis techniques into a multi-modal diagnostic platform. While these technologies still face hurdles to translation into clinical use, they have demonstrated promising applicability towards future pandemics.

An emerging potential application of antibody-based diagnostics is their ability to serve as an at-home monitor of vaccine efficacy. The concept of performing a titer test to check antibodies levels in response to a prior vaccine has been well-established. In fact, titer testing is common for TB[81] and MMR[82]. Given that there are numerous scientific questions regarding both the duration of a SARS-CoV-2 vaccine's protection as well as its universality, discussions are already underway about the possibility of using antibody-based RDTs as a monitoring approach. However, currently, this approach is not recommended by government agencies[83]. One reason is that every RDT targets a different region of the antibody, and therefore, a negative RDT result does not necessarily indicate a low antibody load.

In addition to future pandemics, the spread of multidrug-resistant pathogens means that increased efforts in global health surveillance need to be developed. We can meet this challenge with sustainable, modularly adaptable concepts for POC devices. Powerful capabilities for rapid and comprehensive diagnosis are provided by integration of non-destructive photonic sensors into bioanalytical chip platforms. The antibody-based at-home test kits for COVID-19 coupled with tele-medicine have demonstrated the potential impact that at-home diagnostic assays can have on infectious disease spread.

It is also important to note that this work has focused on discussing strategies to diagnose the onset of infection. While early detection allows for early intervention and isolation, reducing the spread and improving patient outcomes, the ability to accurately monitor illness progression is also important. Initially, hospitals relied solely on RT-PCR for SARS-CoV-2 monitoring which detected viral RNA from both active and inactive infections, resulting in elevated high hospitalization rates[84]. To limit the impact of future pandemics on our healthcare system, we need to develop techniques that can differentiate between active and inactive viral infections.

Therefore, looking into the future, technologists must re-envision where sample collection will occur and the level of expertise of the user. Additionally, miniaturization and automation of systems as well as cost reductions will be important considerations and will play a key role in the eventual translation from the lab to use in patient care. Nevertheless, the COVID-19 pandemic has shown that as diagnostic technologies continue to develop, it is important to both leverage and adapt existing diagnostic platforms to quickly increase accessibility and availability, as well as continue to develop emerging technologies that provide new advantages in sensor sensitivity and specificity.

## AUTHOR INFORMATION

### Corresponding Author

*E-mail: armani@usc.edu

**Author Contributions.**

The manuscript was written through contributions of all authors. All authors have given approval to the final version of the manuscript.

**Notes.**

The authors declare the following competing interests: Jakob R. Jansson, Ian G. de Moura, and Jakub Sram are employees of Fulgent Genetics. All other authors declare no competing interests.

**ACKNOWLEDGMENTS**

This work was supported by the National Science Foundation (2028445), the Federal Ministry of Education and Research (BMBF) in the framework of the projects ReHwIN (13GW0432E), and InfectoXplore (13GW0459A).

Table of contents graphic

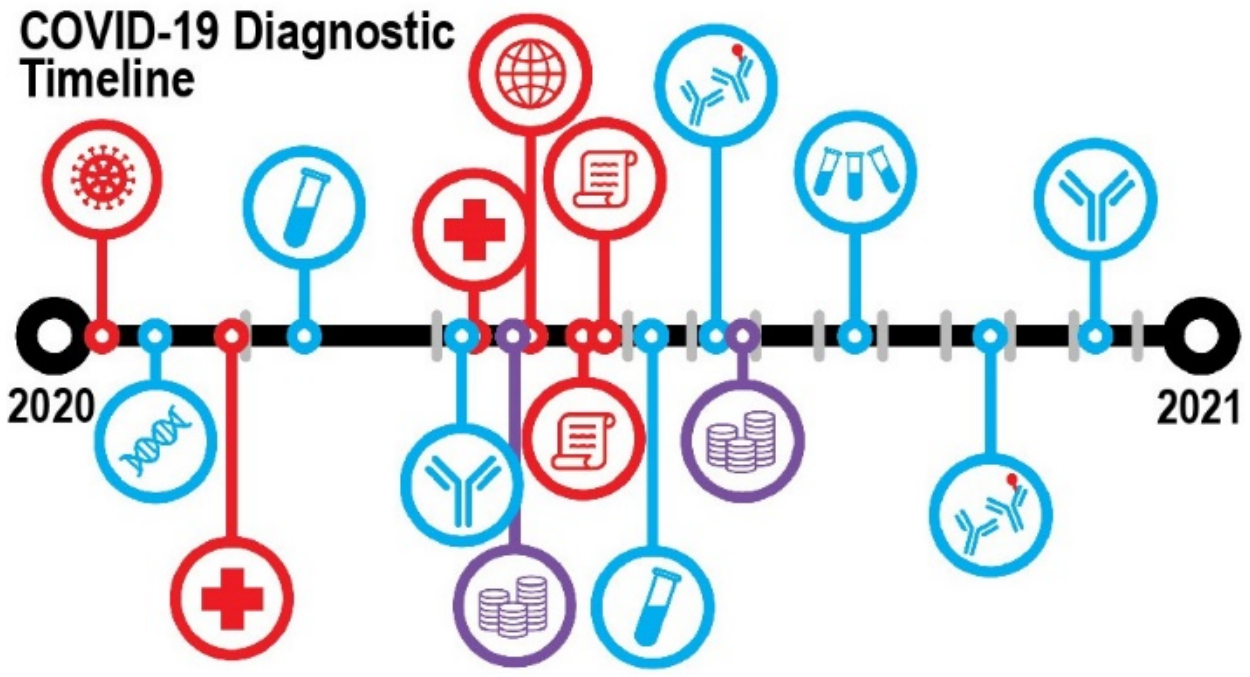